\documentstyle[prl,aps,twocolumn]{revtex}
\begin{document}
\draft
\title{Remark on ``Indication, from Pioneer 10/11, Galileo, and Ulysses 
Data, of an Apparent Anomalous, Weak, Long-Range Acceleration''}
\author{Rainer W. K\"uhne}
\address{Fachbereich Physik, Universit\"at Wuppertal, 42097 Wuppertal, 
 Germany, kuehne@theorie.physik.uni-wuppertal.de}

\maketitle

\pacs{PACS numbers: 04.80.-y, 95.10.Eg, 95.55Pe}

Recently, Anderson et al. \cite{one} presented possible evidence for an 
apparent ``anomalous acceleration'' acting on the spacecrafts Pioneer 
10/11 and, with less statistical significance, also on the spacecrafts 
Galileo and Ulysses.

The anomalous acceleration acting on these spacecrafts is reported to be 
$\sim 8 \times 10^{-8}$cm/s$^{2}$, whereas the upper limit of that acting 
on the orbital motion of the planets Earth and Mars is 
$0.1 \times 10^{-8}$cm/s$^{2}$.

This statement is incomplete, because it gives the (wrong) impression 
that an anomalous acceleration acting on planets and other large objects 
in the solar system is yet unknown. By contrast, the motions of the 
planets Uranus, Neptune, and Pluto show unexplained residuals. 
Furthermore, the motions of a number of comets (including the comets 
Halley, Encke, Giacobini-Zinner, and Borelli) are disturbed by an unknown 
origin.

A number of investigators proposed a non- gravitational force to explain 
the observed anomalous accelerations acting on comets \cite{two}. Such a 
force, however, can hardly be accounted for the spacecraft data, because, 
as recognized by Anderson et al. \cite{one}, it would either violate the 
general relativistic equivalence principle many orders above the 
experimental upper limit or not explain the apparent independence of the 
anomalous acceleration on the distance from the Sun.

Other investigators suggested a transplutonian planet for the explanation 
of the anomalous cometary and planetary accelerations \cite{three}. 
Its mass was predicted to be of the order of that of the planet Saturn. 
However, such a large planet is unlikely to have escaped discovery by 
optical and infrared (especially IRAS) searches.

By contrast, a transneptunian comet or asteroid belt \cite{four} cannot 
yet be excluded. Further examinations are required to show whether this 
hypothetical belt (gravitational effects; resistance by dust particles) 
can be accounted for both the disturbed motions of planets and comets and 
the possible anomalous accelerations acting on spacecrafts.

\end{document}